# Which Information should the UK and US AISI share with an International Network of AISIs?

Opportunities, Risks, and a Tentative Proposal

Lara Thurnherr, King's College London
*(thurnherr.lara@gmail.com)*

**Abstract**

The UK AI Safety Institute (UK AISI) and its parallel organisation in the United States (US AISI) take up a unique position in the recently established International Network of AISIs. Both are in jurisdictions with frontier AI companies and are assuming leading roles in the international conversation on AI Safety. This paper argues that it is in the interest of both institutions to share specific categories of information with the International Network of AISIs, deliberately abstain from sharing others and carefully evaluate sharing some categories on a case by case basis, according to domestic priorities. The paper further proposes a provisional framework with which policymakers and researchers can distinguish between these three cases, taking into account the potential benefits and risks of sharing specific categories of information, ranging from pre-deployment evaluation results to evaluation standards. In an effort to further improve the research on AI policy relevant information sharing decisions, the paper emphasises the importance of continuously monitoring fluctuating factors influencing sharing decisions and a more in-depth analysis of specific policy relevant information categories and additional factors to consider in future research.

**Table of Contents**

# Introduction

The recent establishment of AI Safety Institutes (AISIs) in various countries[1] and US-led inaugural meeting of an international AI Safety Network[2] present an opportunity for enhanced global coordination on AI safety. Participants of the inaugural meeting recognised that grasping this opportunity requires the network to engage in some degree of information sharing and committed to share relevant research findings, results from domestic evaluations, approaches to interpreting tests of advanced systems and technical tools "as appropriate" with each other.[3] AISIs in a jurisdiction with frontier AI companies, like the UK AISI and the US AISI, are in a unique position in this context. With their location in the same jurisdiction as frontier AI companies, they have a more direct access to information. The UK AISI's capacity and comparatively large experience compounds this effect, considering the amount of policy relevant information that is not simply collected from AI companies, but produced (e.g. standards, research methods, etc.).[4] Because of these circumstances, the UK and US AISI have a uniquely strong ability to shape the Network of AISIs access to information.

However, sharing certain kinds of information with an International Network of AISIs couldincrease the probability of leaks and the damage associated with them. There is not always a clear distinction between information that will make AI systems safer and information that will allow AI developers to create more capable systems. Broadening access to the latter could increase the probability of misuse and thus present a risk to international security. It could also empower geopolitical competitors. Safeguarding information that could include intellectual property shared by AI companies is also important, especially while AISIs rely on a collaborative relationship with AI companies for model access and standards development.[5]

This paper explores the potential benefits and risks of the UK and US AISI sharing specific kinds of information with an International Network of AISIs. This work builds on existing literature about AISIs as institutions and the emerging network between them. Adan et al. discuss the structure and operational priorities of the International Network of AISIs.[6] Araujo et al. examine the "first wave" of AISIs, noting common characteristics such as their technical focus, governmental affiliation, and mandates centered on the safety of advanced AI systems.[7] This echoes the results of Ziosi et al. on the roles of AISIs in both domestic and international governance. They emphasize the institutes' involvement in technical evaluations, standardization efforts and international collaboration.[8] Dennis et al. propose a framework to identify AI governance issues that warrant international cooperation. They emphasize four factors determining internationalisation: cross-border externalities, interoperability, uneven governance capacity and risks of regulatory arbitrage.[9]

---

[1] Araujo et al., 2024. More recently, France has announced an AI Safety Institute. (Ministère de l'Économie et des Finances, 2025).
[2] US Department of Commerce, 2024.
[3] NIST, 2024.
[4] Araujo et al.,2024.
[5] Ziosi et al., 2024.
[6] Adan et al.,2024.
[7] Araujo et al., 2024.
[8] Ziosi et al., 2024.
[9] Dennis et al. (2024)

Building on this analysis, the paper then proposes a provisional framework, which could serve as a starting point for decision-making around sharing decisions. It's main function is to distinguish between three categories or tiers of information: "No Brainers", information which would be in the interest of the UK and US AISI to share with the International Network of AISIs, a "Gray Area" which includes information requiring more careful consideration and lastly, "High Trust" Information, which includes information the UK and US AISI would likely not benefit from sharing with the International Network of AISIs. In chapter 4, the paper lists limitations of the framework and areas for future research.

# 1 Benefits of Sharing Information through an AISI Network

This section describes the incentives the UK AISI and US AISI might have to share information with an International Network of AISIs: The benefits of fostering consensus, building trust and credibility and building capacity.

## 1.1 Sharing information to foster consensus

A broader consensus in the International Network of AISIs around AI risk can benefit the sharing party in a number of ways. Firstly, it reduces the burden on domestic frontier AI companies, who might otherwise need to navigate a complex web of divergent AISI expectations across jurisdictions. When AISIs share information to (at least partially) align their evaluation methods or safety standards, companies can focus resources on implementing strong safety measures rather than managing conflicting or widely diverging demands. Secondly, a broader consensus on what reasonable standards might look like could help lower the risk of regulatory arbitrage - a "race to the bottom" where AI developers might relocate to those jurisdictions with the weakest safety standards. Considering the transnational nature of some risks from AI, lowering the risk of "regulatory havens" would be in the interest of all members of the international community.[10] By sharing safety standards, risk assessment methods and supplying supporting empirical evidence for these standards, the UK AISI and US AISI could contribute to this endeavour.

## 1.2 Building Trust

As AISIs in countries with frontier AI companies, the UK AISI and US AISI are in a privileged position, with more direct access to policy relevant information. But as frontier model capabilities and potential global risks from misuse, accidents or loss of control rise, these institutions could also be perceived to face a greater responsibility of preventing risks.

A commitment to transparency could serve as a "confidence building measure" and a signal that the UK and US AISI are taking their responsibility seriously. The trust from this perception could increase these countries' credibility and the likelihood that the participants of the International Network of AISIs will cooperate effectively in the future. Establishing trust and credibility in practically oriented forums like the one in question could also lay the foundation for more advanced collaboration in the future, should this become necessary.[11]

---

[10] ibid.
[11] Zelikow et al. 2024.

## 1.3 Building Capacity

Both in the UK and in the US, AISI capacity is limited. Sharing information could play a key role in reaching their goals sooner and more efficiently by broadening the number of international partners who could contribute to common goals. If a larger number of AISIs have early access to preliminary safety research results, research agendas on risk managements, evaluation methods, and, potentially, software infrastructure[12] from the UK and US AISI, they could avoid unnecessarily duplicating work and are thus more likely to contribute to the frontier of these fields. This could also contribute to more informed productive discussions at network meetings and increase the trustworthiness of the sharing parties.

**Fig. 1) Benefits of Sharing Information with an International Network of AISIs**

| What? | **Building Capacity** | **Building Trust** | **Fostering Consensus** |
|---|---|---|---|
| Why? | - Reach common (research) goals faster and more efficiently<br>- Potentially increase interoperability | - Lower risk of potential backlash or conflict<br>- Increase credibility of research and proposals | - Lower risk of a "race to the bottom"<br>- Align research priorities and risk prevention policies |
| How?*<br><br>*By sharing which info? | Evaluation tools | Data on current AI systems (e.g. Evaluation Results) | |
| | Open questions or research agendas | Forecasts on future AI systems and potential transnational risks | |
| | Research- or Evaluation Methods | Proposals for standards or risk prevention measures | |
| | Institutional learnings (e.g. "How to scale an AISI") | | |

## 2 Risks of Sharing Information

### 2.1 Oversharing Proprietary Information and Losing Trust from AI Companies

Oversharing some information could lead to a loss of trust from AI companies, which have strong commercial incentives to protect their intellectual property. If the UK or US AISI mishandle this information or share it too widely, it could result in several negative consequences.

Firstly, AI companies might withhold information from their respective AISI if they are not legally compelled to share it, which could be detrimental to the AISI's ability to fulfil its mandate, especially if it lacks the authority to enforce information sharing or doesn't know what specific information to request. Secondly, collaboration on standards and best practices would likely become more challenging if the relationship between an AISI and AI companies turns more adversarial.

---

[12] See UK AISI, 2024.

## 2.2 Risks to National Security and AI Leadership from Leaks

Sharing information with a greater number of actors increases the probability of a leak. If advanced capabilities or crucial intellectual property, such as model weights, fall into the hands of competing state actors, they could catch up more quickly to leading domestic AI companies.[13] If they fall into the hands of malicious non-state actors, risks from misuse (and potentially accidents)[14] rise. What constitutes "capabilities research" isn't always obvious: Many evaluation methods require specific methods to elicit dangerous capabilities, and sharing these methods could aid in the misuse or dangerous applications of existing or future AI models.[15]

Leaked evaluation results might have similar effects because they can guide competing or malicious actors toward research directions they wouldn't have considered promising. Furthermore, research into making AI systems safer also makes them more reliable. This makes the distinction between "safety-focused research" and "capabilities-focused research" conceptually challenging and complicates the question of what to share.

## 2.3 Miscellaneous other Potential Challenges

Apart from the risks outlined above, the following three potential challenges from sharing information are also worth noting. While they are not addressed in the framework used in this paper, future research could benefit from an in-depth inquiry of their likelihood and potential consequences.

1. AISIs wasting time or capacity: Acquiring specific information, correctly identifying which information to share and then sharing information can require time and money. An expectation to share information which doesn't exist yet requires additional resources.
2. Research "monoculture": Duplication in nascent fields can be important. Some research results and methodologies should be scrutinised by multiple AISIs. Sharing too much information or avoiding duplication completely might lead to a research monoculture, where all "eggs are in one basket".
3. Lock-in effects: Once an AISI shares a strategy or research agenda with several actors, it might be harder to correct a strategy later on.

---

[13] Nevo et al., 2024.
[14] Assuming that malicious non-state actors are unlikely to follow safety protocols for model development & deployment.
[15] Phuong et al., 2024.

# 3 Information Sharing Framework: Categorisation and Examples

To take the risks of sharing information into account when deciding which information (not) to share, this paper proposes a red-amber-green analysis of whether the information exhibits characteristics which could increase the risks outlined in chapter 2.2. As visualised in Fig. 2), if we assume a hypothetical piece of information A could be beneficial to share, this paper proposes analysing the potential effects of sharing this information through assessing the proprietary nature of the information and its utility for capability advances or misuse if leaked.

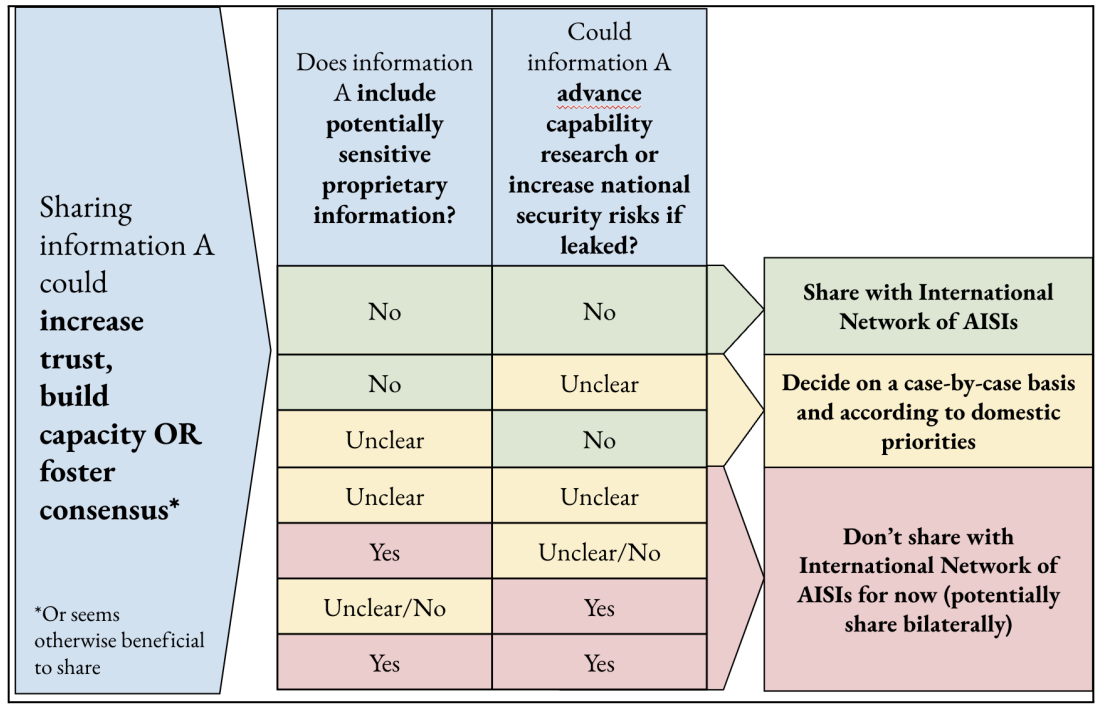

**Fig. 2) Provisional Framework: Taking the Risks of Sharing "Information A" into Account**

After this process, three categories emerge:

1) **"No Brainers"**: Low-risk information that should be shared by default with
2) **A "Gray Area"**: Information requiring careful consideration and sharing depends on a range of priorities.
3) **"High Trust" Information**: Sensitive information, not be shared by default with the International Network of AISIs (but might still be useful to share bilaterally with specific trusted allies).

The following three sections provide preliminary results from an application of this framework.

## 3.1 "No Brainers"

Sharing **Institutional learnings and strategies** (e.g. learnings on how to scale an AISI, which institutional norms and organisational structures improve research results and speed) could increase the capacity at other AISIs and signal a commitment to common interests. This could also include future research agendas and open questions, which could decrease unnecessary duplication.

Table 1) "No Brainers": Risks, Benefits and Uncertainties

| Information Category | Trust from Companies | AI Leadership and Security | Builds Consensus | Builds Trust | Builds Capacity |
|---|---|---|---|---|---|
| Institutional Learnings and Strategy | ✓ | ✓ | ⚖ | ✓ | ✓ |
| Evaluation Standards | ✓ | ✓ | ✓ | ⚖ | ✓ |
| Risk Management Standards | ✓ | ✓ | ✓ | ✓ | ⚖ |
| Evaluation Infrastructure | ✓ | ✓ | ⚖ | ⚖ | ✓ |
| AI Incident Definitions | ✓ | ✓ | ✓ | ✓ | ✓ |

(✓ = Low Risk/High Benefit, ⚖ = Unclear Risk/Benefit)

Sharing **risk management standards** (e.g. standards on how different risks from AI should be estimated) could spur a similar international consensus building process. While setting risk thresholds might be a political decision which each jurisdiction might rightly want to decide on its own, a consensus on the likelihood and implications of a particular risk event should be grounded in evidence based discussions, to which sharing AISIs can contribute. These standards can also communicate effectively how seriously the UK and US AISI takes transnational risk events and thus foster trust.

**AI incident definitions** (e.g. what should be classified as an incident? How much financial or physical damage is required for such a classification? What does it mean for an AI system to "cause" an incident) are an example of a specific open question which might become important to clarify for future domestic and international policy initiatives. Discussing these definitions would not require the UK or US AISI to share them with an international network, but it could make the potential establishment of an international incident reporting system easier, avoid misunderstandings on such questions, signal a commitment to the safety of the international community and guide domestic and foreign AISI research in a productive direction. **Sharing evaluation standards** (e.g. which models should be tested? how should they be evaluated?) could contribute to a more productive conversation on how evaluations should be developed, conducted and interpreted. Given the significance of evaluations to the mandates of many AISIs[16] and the nascent state of the practice,[17] this could help significantly improve AISIs capacity to develop, conduct or interpret evaluations.

Sharing **evaluation infrastructure** like the UK AISIs "inspect" platform could improve the speed at which foreign AISIs can contribute to a global evaluations ecosystem.

---

[16] Araujo et al, 2024.
[17] Apollo Research, 2024.

3.2 The "Gray Area"

Sharing unpolished **preliminary safety and evaluation research** results on problems other AISIs could be working on could help avoid unnecessary duplication and build trust, as well as establish a higher level of consensus around the current state of safety and evaluation research. It could also, in some cases, increase the risk of leaking capability research if not examined thoughtfully.

Sharing **AI risk estimates and capability forecasts** (e.g. estimates on which capabilities future models will exhibit and corresponding risks that could arise as a consequence) could improve the foresight of and consensus between different jurisdictions. However, if risk estimates for specific misuse fall into the hands of a malicious actor, that knowledge may become an incentive for them to gain unauthorised access to a model.

**Table 2) The "Gray Area": Risks, Benefits and Uncertainties**

| Information Category | Trust from Companies | AI Leadership and Security | Builds Consensus | Builds Trust | Builds Capacity |
|---|---|---|---|---|---|
| Preliminary Safety and Evaluation Research | ✓ | ⚖ | ✓ | ⚖ | ✓ |
| Pre-Deployment Evaluation Results (shared after deployment) | ✓ | ⚖ | ✓ | ✓ | ⚖ |
| AI Risk Estimates and Capability Forecasts | ✓ | ⚖ | ✓ | ✓ | ✓ |
| Post-Deployment Evaluation Results | ✓ | ⚖ | ✓ | ⚖ | ⚖ |
| Detailed Evaluation Methods | ✓ | ⚖ | ⚖ | ⚖ | ✓ |

(✓ = Low Risk/High Benefit, ⚖ = Unclear Risk/Benefit)

Sharing **detailed evaluation methods** from start to finish, potentially including tested tasks and methods of dangerous capability elicitation. While sharing these methods would be helpful to increase the number of AISIs who can perform evaluations, leaks of these methods could help malicious actors elicit the dangerous capabilities these evaluations bring attention to.

The level of detail at which **pre-deployment evaluation results (after** deployment) are shared matters: While sharing detailed evaluation results would undoubtedly increase trust among AISI's and consensus around the current frontier of AI models, sharing some results unilaterally could negatively affect the relationship between developers and the UK or US AISI, even if the sharing after deployment likely reduces these cases. The same considerations apply to **post-deployment evaluation results**.

## 3.3 High Trust Information

**Sharing private sector cyber incident reports** on incidents at major frontier AI companies could, if the reports are detailed enough, lead to broader knowledge and potential leaks of the vulnerabilities of frontier companies, increase the risk of model theft. While a commitment to sharing reports on these incidents could establish a higher level of trust and prepare other AISIs early for the effects of potential misuse of a stolen frontier model, the International Network of AISIs, in its current state, doesn't seem to be the optimal forum for sharing information this sensitive. (Although, again, sharing less detailed reports could potentially be shared).

Sharing **pre-deployment evaluation results** shared **before** deployment.could damage the relationship between the AISI and AI companies, because of the increased risk that valuable information might leak to economic competitors. They could also increase the risk of geopolitical competitors using this information to endanger the technological advantage of the UK or the US. The same considerations apply to **pre-deployment model access** and **sensitive proprietary model information** (i.e. Model weights, proprietary datasets, algorithmic secrets, or extensive development process knowledge).

**Table 3) "High Trust": Risks, Benefits and Uncertainties**

| Information Category | Trust from Companies | AI Leadership and Security | Builds Consensus | Builds Trust | Builds Capacity |
|---|---|---|---|---|---|
| Private Sector Cyber Incidents | ⚖ | ⚖ | ✓ | ✓ | ⚖ |
| Pre-Deployment Evaluation Results (Shared before deployment) | ⚖ | ⚖ | ✓ | ✓ | ⚖ |
| Pre-deployment Model Access | 🛡 | ⚖ | ⚖ | ✓ | ✓ |
| Proprietary Model Information | 🛡 | 🛡 | 🛡 | ⚖ | 🛡 |

(✓ = Low Risk/High Benefit, ⚖ = Unclear Risk/Benefit, 🛡 = High Risk/Low Benefit )

## 4 Limitations and Future Research Directions

The findings in this paper depend on the likely temporary relevance of a variety of factors: The UK and US AISI might lose their information advantage, the International Network of AISIs might change in structure, current participants in the network might withdraw from the network and new AISIs might join. The strategic value of maintaining a cooperative relationship between AI companies and AISIs may diminish, if AISIs in the UK and the US get an increased power to compel companies to share information with them. Because of this, this framework will require frequent adaptation.

The framework presented in this paper is applied to generalised information categories. All of them can and should be split up into more fine-grained subcategories in future research on this topic. This would reduce the number of "unclear" designations of risks and benefits and enable more actionable policy recommendations. For example, evaluation results on a particular topic might be much more sensitive than others, AI risk estimates and capability forecasts might produce a larger risk to the competitive advantage of sharing parties if they are shared at a specific level of detail. More conceptual or technical work could address the different dimensions of "dual-use" information, identifying definitions and scenarios under which information could result in both capability advances and enhanced safety.

Future research could additionally take into account Dennis et al. 's identification of the importance of interoperability in internationalised information.[18] Incorporating this factor into the framework, either as a component of the capacity benefit or as a separate benefit to consider could be helpful. Lastly, A closer examination of the miscellaneous other challenges alluded to in chapter 2.3 (Sharing AISIs wasting time or capacity, accidentally creating a research "monoculture" and potential lock-in effects through sharing) could help expand this framework to include a more comprehensive view.

# 5 Conclusion

This paper has proposed a framework for information sharing between AI Safety Institutes, with a particular focus on AISIs in jurisdictions with frontier AI companies sharing information with an International Network of AISIs. The three-tier framework—distinguishing between information that should be shared by default ("No Brainers"), information requiring case-by-case assessment ("Gray Area"), and information needing strong trusting relationships ("High Trust")—provides a starting point for AISIs in jurisdictions with frontier AI companies to make complex sharing decisions. This framework recognizes that while some information sharing is essential for building consensus and capacity across jurisdictions, other types of information sharing could potentially increase risks through capability leaks or damaged industry relationships.

# 6 Acknowledgments

This research was conducted during the 2024 Summer Fellowship at the Centre for the Governance of AI under the supervision of Helen Toner and research management of Valerie Belu, both of whom I'd like to thank sincerely for their feedback and help during the research process. Furthermore, I am grateful for the feedback of and discussions with Renan Araujo, Oliver Guest, Nikhil Mulani, Arthur Goemans, Claire Dennis, Michel Justen, Lucia Velasco, James Petrie, Tegan McCaslin, Charles Martinet, Casey Mahoney, Markus Anderljung, Joe Castellano, Robert Trager, Neel Alex, Shira Gur-Arieh, Harrison Durland, Cassidy Bereskin, Sam Manning and other members of the community of researchers and practitioners in and around the AISI ecosystem and the field of international AI governance.

---

[18] Dennis et al., 2024.

# 7 Sources

Zelikow, Philip, Cuellar, Mariano-Florentino (Tino) Cuéllar, Schmidt Eric and Matheny, Jason: Defense Against the AI Dark Arts: Threat Assessment and Coalition Defense, 2024. Hoover Institution. (Available at: https://carnegieendowment.org/research/2024/12/defense-against-the-ai-dark-arts-threat-assessment-and-coalition-defense?lang=en¢er=europe).

Ziosi, Marta, Dennis, Claire, Trager, Robert, Bucknall, Ben, Campos, Simeon, Martinet, Charles, Smith, Adam and Stein, Merlin: AISIs' Roles in Domestic and International Governance. 2024. (Available at: https://www.oxfordmartin.ox.ac.uk/publications/aisis-roles-in-domestic-and-international-governance).
12